\documentclass[prb,preprint,showkeys,preprintnumbers,amsmath,amssymb]{revtex4-1}

\usepackage{amsmath,amssymb,bm,epsfig}
\usepackage{dcolumn}
\usepackage{bm}
\usepackage{feynmf}

\begin{document}

\preprint{to appear in Phys. Rev. B}

\title{Comment on ``Theory of Phonon-Assisted Adsorption in Graphene: Many-Body Infrared Dynamics''}

\author{Dennis P. Clougherty}

\affiliation{
JILA, National Institute of Standards and Technology and University of Colorado, 440 UCB, Boulder, CO 80309}

\affiliation{
Department of Physics, 
University of Vermont, 
Burlington, VT 05405-0125}

\date{\today}

\begin{abstract}
Two approximations used by Sengupta [Phys. Rev. B {\bf 100}, 075429 (2019)] in numerically computing the adsorption rate of cold hydrogen atoms on suspended graphene are critically examined.  The independent boson model approximation (IBMA) was used to compute the atom self-energy, and the single-pole approximation (SPA) was used to obtain the adsorption rate from the self-energy.   It is shown explicitly that there are additional contributions to the self-energy appearing at the same order of the atom-phonon coupling as the IBMA terms that alter the value of the real part of the self-energy at low energies by several orders of magnitude in the regime of interest.   This shift in the self-energy consequently renders the use of SPA invalid.

\end{abstract}

\maketitle
The work of Ref.~\onlinecite{Sengupta2019} revisits a physisorption model proposed by Clougherty \cite{DPC2013} to describe the effects from the infrared phonon dynamics on sticking to 2D materials.  This work relies on the assumption that the self-energy of the slow incident atom can be approximated by the 1-loop diagram where the atom propagator is replaced by the propagator of the independent boson model (IBM), an approximation henceforth referred to as the independent boson model approximation (IBMA).  This approximation was not derived; consequently, the regime of validity for the IBMA was not obtained.

Using the IBMA self-energy,  the sticking rate is calculated numerically by evaluating the imaginary part of the IBMA self-energy at the atom energy $E_k$.  This single-pole approximation (SPA) to the real-time atom Green's function rests on a number of assumptions about the behavior of the self-energy \cite{mahan}.  Additionally, Ref.~\onlinecite{Sengupta2019} finds the quasiparticle weight as $Z\approx 0.99$, for all atom energies $E_k$ considered.

From numerical results based on these approximations, Ref.~\onlinecite{Sengupta2019} concludes that for suspended micromembranes of graphene at $T=10$ K, the adsorption rate will be finite, independent of the membrane size, and in agreement with the lowest-order perturbative result obtained by Fermi's golden rule.  This conclusion is in disagreement with two previous theoretical studies that concluded, at low atom energies, the sticking rate is severely suppressed by a phonon orthogonality catastrophe \cite{dpc10,dpc11,DPC2013, dpcSPR2017}.

In this Comment, the two approximations of Ref.~\onlinecite{Sengupta2019} are critically examined.   To assess the validity of the IBMA, the exact closed-form expression for the atom self-energy to quadratic order in the atom-phonon coupling ($O(g_{kb}^2)$) is obtained.   This order $O(g_{kb}^2)$ self-energy includes many contributions neglected in the IBMA.  This result is then compared to the IBMA self-energy for parameter values used in Ref.~\onlinecite{Sengupta2019}.  Finally, the $O(g_{kb}^2)$ self-energy is  used to examine the validity of SPA in approximating the adsorption rate.   

Using the model of Eqs.~[1-3] in Ref.~\onlinecite{Sengupta2019}, the exact atom self-energy to order $O(g_{kb}^2)$ is 
\begin{equation}
\Sigma(t)=-i g_{kb}^2 \sum_{m,n} \langle T[X(t)b(t)A_m(t)X^\dagger(0)b^\dagger(0)A_n(0)]\rangle_\beta
\label{se}
\end{equation}
 where $A_n=a_n+a_n^\dagger+{2 \lambda_n}b^\dagger b$ and $X=\exp(\sum_p \lambda_p(a_p^\dagger-a_p))$ and $\lambda_p=g_{bb}/\omega_p$.
 Here,  $\langle \dots \rangle_\beta={\cal Z}^{-1} {\rm Tr} (e^{-\beta H_{ph}} \dots)$ with $H_{ph}= \sum_n \omega_n a^\dagger_n a_n$.  ${\cal Z}$ is the phonon partition function.

 \begin{figure}
\includegraphics[width=12cm]{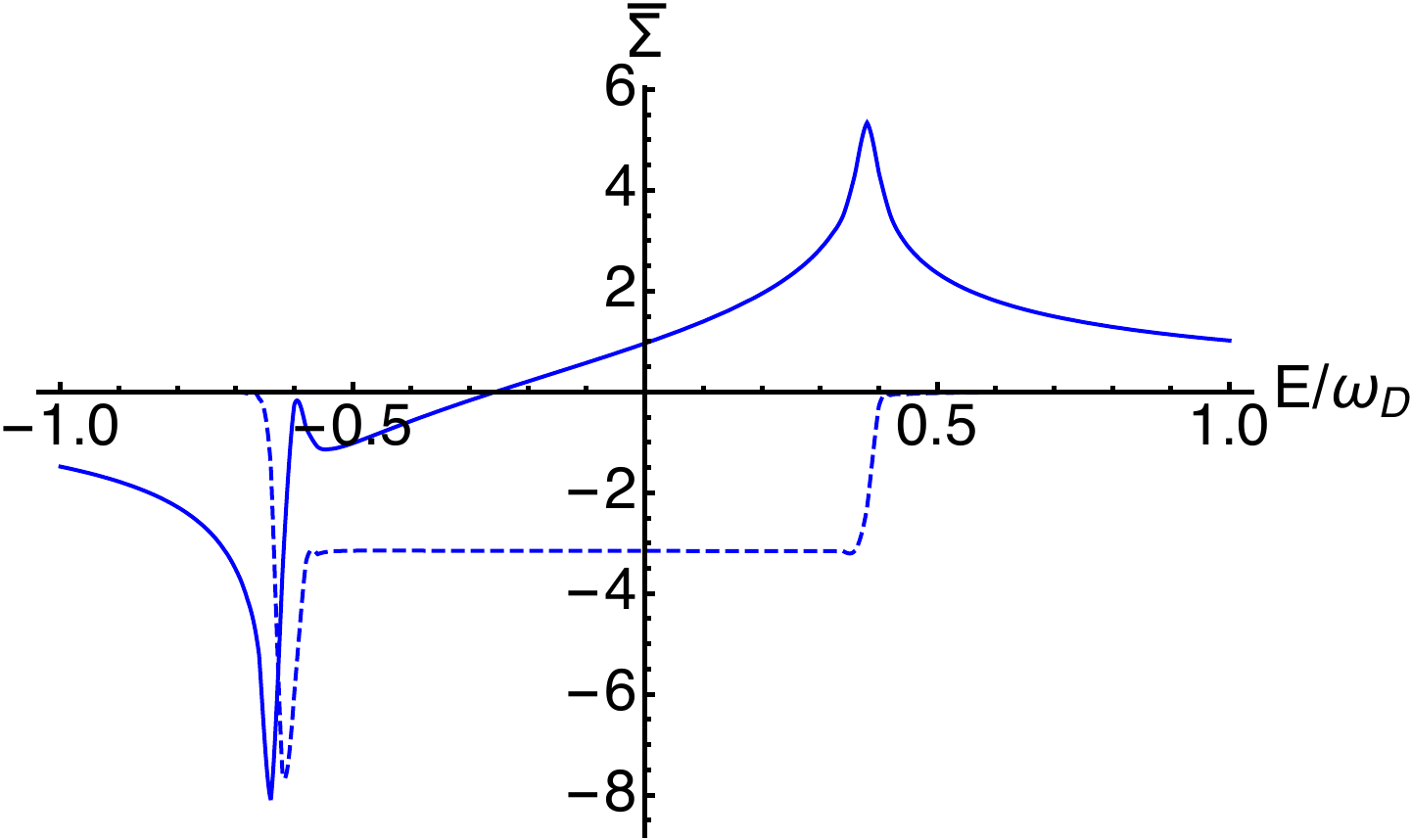}
\caption{\label{fig:IBMAse} Real (solid) and imaginary (dotted) parts of the atom self-energy in the IBMA $\bar\Sigma^{(\rm IBM)}$ versus $E/\omega_D$ for $\epsilon=2.0$ K.  The (dimensionless) self-energy is defined $\bar\Sigma\equiv\Sigma/g_{kb}^2\rho_0$ where $\rho_0$ is the partial (axisymmetric) vibrational density of states for the membrane.}
\end{figure}
 
The IBMA self-energy $\Sigma^{(\rm IBM)}(t)$ can be obtained from Eq.~\ref{se} by factorizing the matrix element as
 \begin{equation}
  \langle T[X(t)b(t)A_m(t)X^\dagger(0)b^\dagger(0)A_n(0)]\rangle_\beta\to  \langle T[X(t)X^\dagger(0)]\rangle_\beta  \langle T[b(t)b^\dagger(0)]\rangle_\beta  \langle T[A_m(t)A_n(0)]\rangle_\beta
 \end{equation}
 The product of the first two factors is recognized as the IBM Green function.
Thus, this factorization gives a result equivalent to the replacement of the bound atom Green function by the IBM Green function in the 1-loop self-energy.  The IBMA self-energy is given by (Eq.~6 in Ref.~\onlinecite{Sengupta2019})
\begin{equation}
\Sigma^{(\rm IBM)}(E)= g_{kb}^{2} \sum_{q}\bigg(n_{q} G^{\rm IBM} (E+\omega_{q})+ (n_{q} +1) G^{\rm IBM} (E-\omega_{q})\bigg)\\
\label{IBMA}
\end{equation}
where $n_q$ is the average number of phonons with $\omega_q$ with the membrane at temperature $T$ and $G^{\rm IBM}$ is the bound atom Green's function from the independent boson model (Eqs.~(28) and (29) in Ref.~\onlinecite{Sengupta2019}).   A plot of the real and imaginary parts of $\Sigma^{(\rm IBM)}$ as a function of the energy $E$ are shown in Fig.~\ref{fig:IBMAse}.

\begin{figure}
\includegraphics[width=16cm]{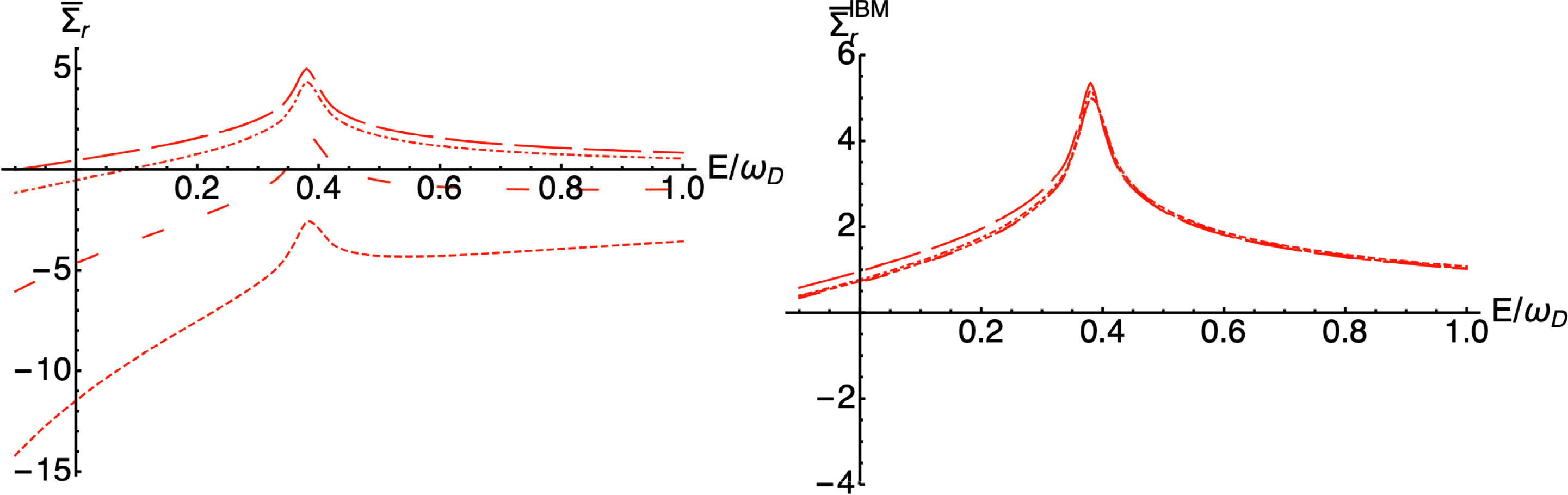}
\caption{\label{fig:SRcompare1} Real part of the (scaled) atom self-energy $\bar\Sigma_{r}$ versus $E/\omega_D$. IBMA  (right) compares favorably to the $O(g_{kb}^2)$ (left) for $\epsilon=2$ K (long dashed).  The real part of the $O(g_{kb}^2)$ self energy shifts downward as $\epsilon$ is reduced: $\epsilon=$1.5 (dot-dashed), 1 (dashed), and $0.8$ K (short dashed). Order $O(g_{kb}^2)$ self-energy (left) changes substantially over the range of $\epsilon$.  IBMA self-energy (right) changes little with $\epsilon$ over the same range.}
\end{figure}

There are, however, additional terms in the self-energy, beyond the IBMA terms.  The most IR singular of these terms result from the noncommutativity of  the displacement operator $X$ and the phonon operator $A_n$, as $[A_n,X]=2\lambda_n X$.  Furthermore, Wick's theorem and diagrammatic expansions based on this theorem cannot be used, as the commutator is not a  simple c-number \cite{fetter}.

\begin{figure}
\includegraphics[width=12cm]{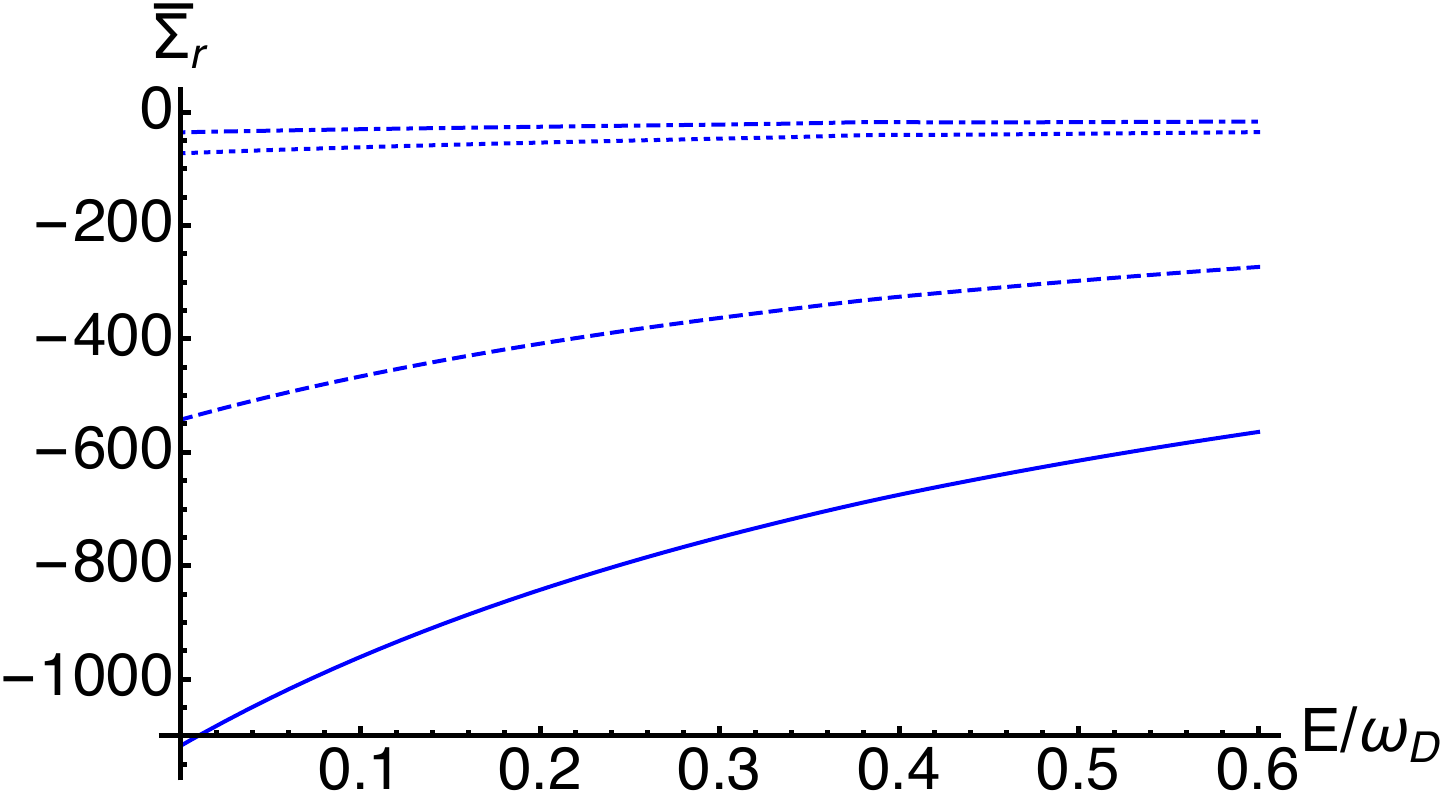}
\caption{\label{fig:SRcompare2} Real part of the (scaled) atom self-energy $\bar\Sigma_{r}$  to order $O(g_{kb}^2)$  versus $E/\omega_D$ for $\epsilon=0.25$ (solid), 0.3 (dashed), 0.5 (dotted), and $0.6$ K (dot-dashed).  }
\end{figure}

Disentangling the matrix element using the commutation relations yields the exact time-ordered finite temperature self-energy to order $O(g_{kb}^2)$ (and to all orders in $g_{bb}$).  Its Fourier transform $\Sigma(E)$ is found to be 
\begin{equation}
\Sigma(E)=\Sigma^{(a)}(E)+\Sigma^{(b)}(E)
\end{equation}
where
\begin{eqnarray}
\Sigma^{(a)}(E)&=& g_{kb}^{2} \sum_{q}\bigg((2\Lambda\lambda_q-2n_q^2\lambda_q^2)G^{\rm IBM} (E)+(n_{q}(2\Lambda\lambda_q+1)+2n_q^2\lambda_q^2)
G^{\rm IBM} (E+\omega_{q})\nonumber\\
&+&((n_{q} +1)(1-2\Lambda\lambda_q)+2n_q^2\lambda_q^2) G^{\rm IBM} (E-\omega_{q})+n_q\lambda_q^2(1-n_q)G^{\rm IBM} (E+2\omega_{q})\nonumber\\
&-&(n_q+1)\lambda_q^2 n_qG^{\rm IBM} (E-2\omega_{q})\bigg)
\label{Og2a}
\end{eqnarray}

\begin{eqnarray}
\Sigma^{(b)}(E)&=& g_{kb}^{2} \sum_{p, q}\bigg(-\lambda_p\lambda_q(1+2n_q n_p+n_q+n_p)G^{\rm IBM} (E)+n_q n_p\lambda_q\lambda_p G^{\rm IBM} (E+\omega_{q}+\omega_p)\nonumber\\
&+&(n_{q} +1)(n_{p} +1)\lambda_q\lambda_p G^{\rm IBM} (E-\omega_{q}-\omega_p)-(n_q+1)n_p\lambda_q\lambda_p G^{\rm IBM} (E-\omega_{q}+\omega_p)\nonumber\\
&-&(n_p+1) n_q\lambda_q\lambda_p G^{\rm IBM} (E+\omega_{q}-\omega_p)\bigg)
\label{Og2b}
\end{eqnarray}
and $\Lambda=\sum_p \lambda_p$.

\begin{figure}
\includegraphics[width=16cm]{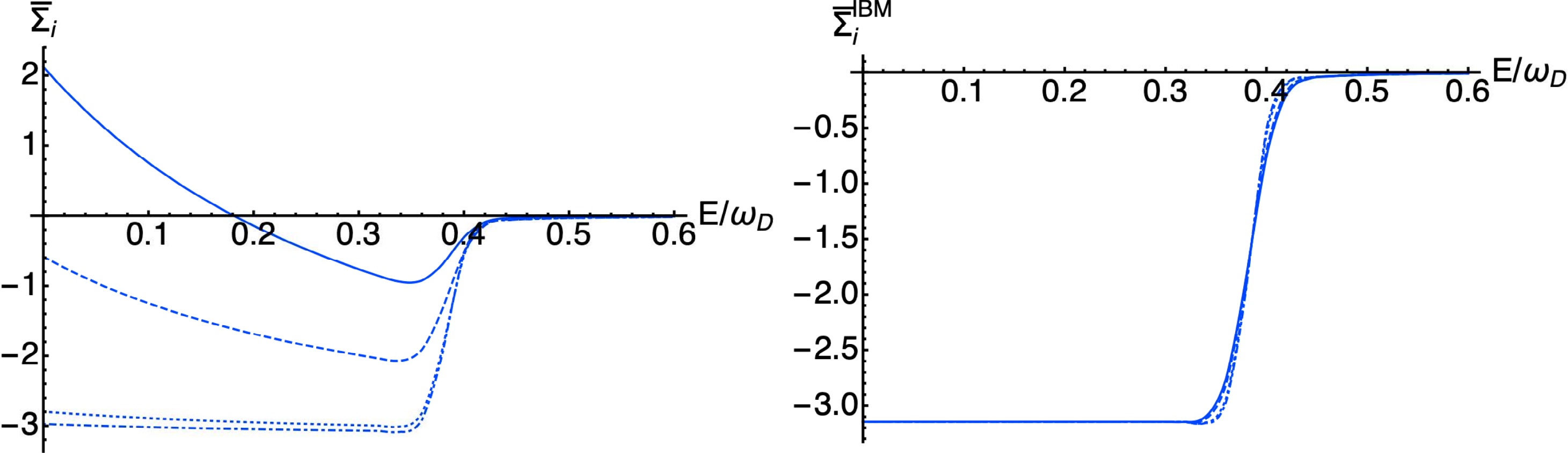}
\caption{\label{fig:SIcompare} Imaginary part of the atom self-energy $\bar\Sigma_i$ versus $E/\omega_D$.   Order $O(g_{kb}^2)$ self-energy $\bar\Sigma_i$ (left) becomes positive near $E=0$ with decreasing $\epsilon$, while IBMA self-energy $\bar\Sigma_i^{\rm IBM}$ changes little for $\epsilon=0.25$ (solid), 0.3 (dashed), 0.5 (dotted), and $0.6$ K (dot-dashed).}
\end{figure}

It is noted that the IBMA contributions of Ref.~\onlinecite{Sengupta2019} are contained in Eq.~\ref{Og2a}; however,
the additional terms beyond the IBMA dominate the self-energy at low energies for sufficiently small IR cutoff $\epsilon$.  While the IBMA results are in agreement with the naive Golden rule results ($g_{bb}=0$) over the range of IR cutoffs considered in the numerical work of Ref.~\onlinecite{Sengupta2019},  the inclusion of the additional contributions contained in the $O(g_{kb}^2)$ self-energy can shift the real part of the self-energy by orders of magnitude near $E=0$ (see Fig.~\ref{fig:SRcompare2}).  

\begin{figure}
\includegraphics[width=12cm]{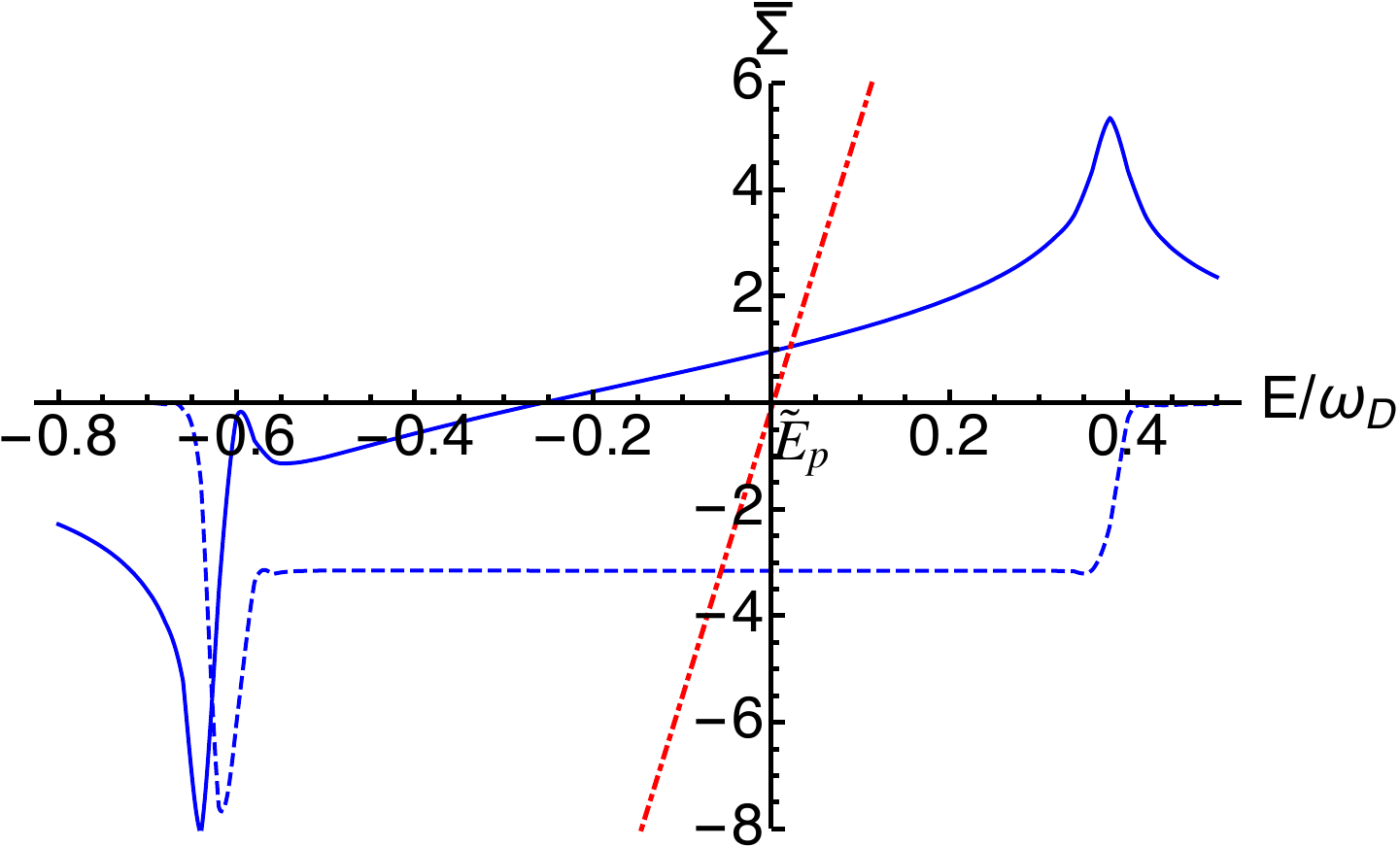}
\caption{\label{fig:QP4} Graphical solution for the quasiparticle energy $E_p$.  Real part (solid) of the atom self-energy in IBMA ${\rm Re}\ \bar\Sigma^{({\rm IBM})}$  for $\epsilon=2.0$ K versus $ E/\omega_D$ crosses the line (dot-dashed) where ${\bar E}_p-{{\bar E}}_k={\rm Re}\ \bar\Sigma({{\bar E}_p})$.  In the single-pole approximation, the adsorption rate depends on ${\rm Im}\ \Sigma^{({\rm IBM})}(E_p)$ and is determined by Eq.~\ref{rate}.}
\end{figure}

Figure \ref{fig:SRcompare1} illustrates that for $\epsilon=2$ K, the IBMA and the exact order $O(g_{kb}^2)$ self-energy are in good agreement over a substantial range of energies $E$.  However, as $\epsilon$ is decreased, the real part of the self-energy $\Sigma_{r}$ is shifted downward relative to $\Sigma_{r}^{(\rm IBM)}$ near $E=0$.  

There are also substantial changes to the imaginary part of the self-energy from the additional terms.
Figure \ref{fig:SIcompare} shows that $\Sigma_{i}$  shifts upward near $E=0$ with decreasing $\epsilon$, while $\Sigma_{i}^{(\rm IBM)}$ is insensitive to changes in $\epsilon$.

\begin{figure}
\includegraphics[width=12cm]{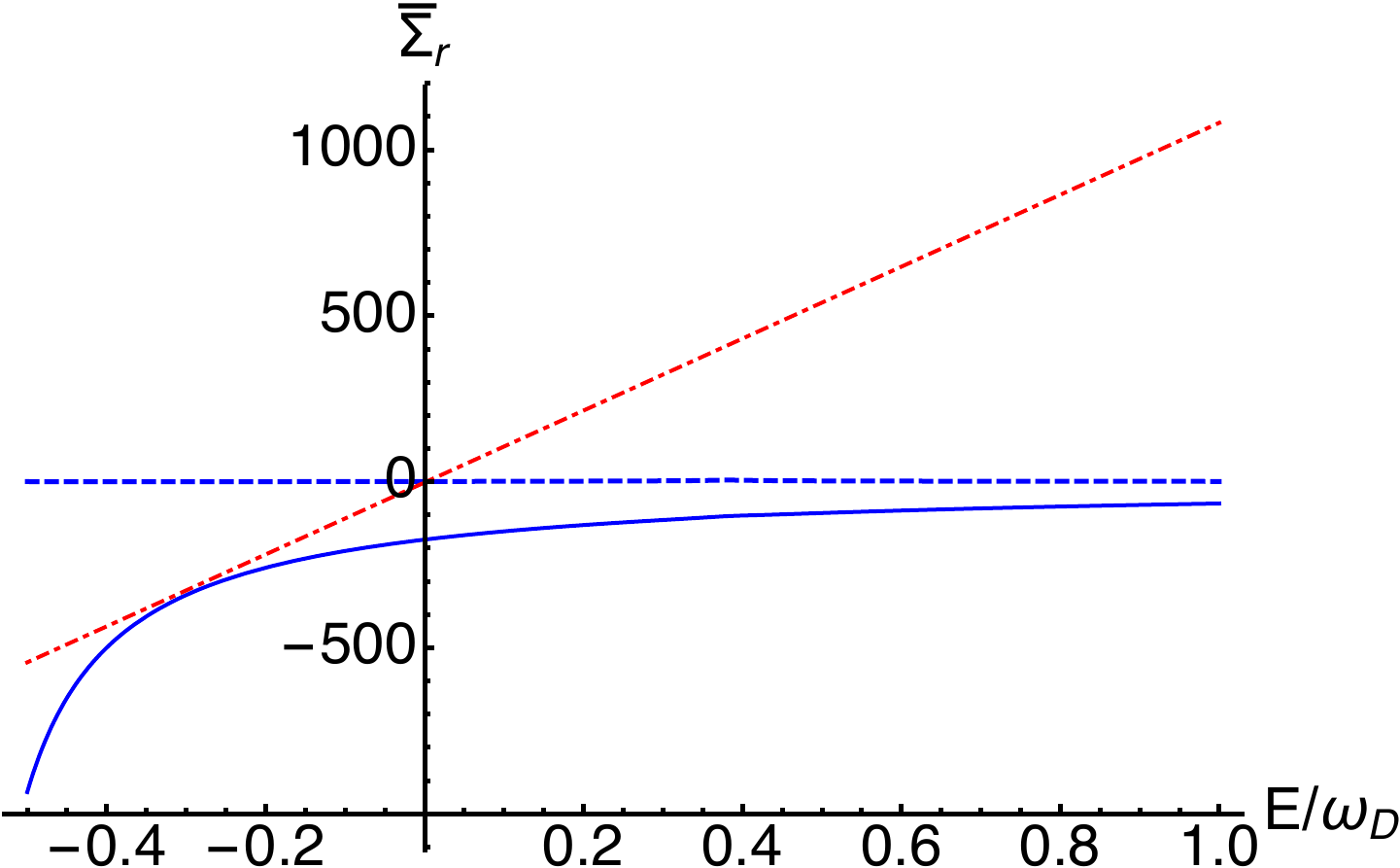}
\caption{\label{fig:QP} Graphical solution for the quasiparticle energy $E_p$ fails with large negative shifts of the real part of the self energy.  Real part (solid) of the atom self-energy ${\rm Re}\ \bar\Sigma$ for $\epsilon\le 0.4$ K versus $E/\omega_D$ does not intersect with the line (dot-dashed) where ${\bar E}_p-{{\bar E}}_k={\rm Re}\ \bar\Sigma({{\bar E}_p})$. The single-pole approximation fails for low atom energy ($E_k\ll \Sigma_r(0)$)  when using the order $O(g_{kb}^2)$ self energy, but does not using the IBMA self energy (horizontal dashed) line.}
\end{figure}

The real part of the self-energy $\Sigma_{r}$ at $E=0$ is found to be negative for the values of $\epsilon \lesssim 1.5$ K, while $\Sigma_r^{\rm IBM}(0)$ is positive over the range of $\epsilon$ considered in Ref.~\onlinecite{Sengupta2019}.  Analytically constructing the asymptotic expansion of Eq.~\ref{IBMA} for $\epsilon\to 0$, it is seen that $\Sigma_r^{\rm IBM}(0)$ diverges slowly (logarithmically) in the  limit of $\epsilon\to 0$, a result that is not readily apparent for $\epsilon \gtrsim 0.1$ K but becomes clear with further decreases in $\epsilon$.  This result has been confirmed by  numerical calculations.  In contrast, the real part of the order $O(g_{kb}^2)$ self-energy  at $E=0$, $\Sigma_{r}(0)$, diverges more rapidly (algebraically) as $\epsilon\to 0$.  This rapid growth of $\Sigma_{r}(0)$ with decreasing $\epsilon$ is clearly visible for $\epsilon\lesssim 0.5$ K as seen in Fig.~\ref{fig:SRcompare2}.

 It is also noted that the curvature of the real self-energies  differ in sign, with the real part of exact order $O(g_{kb}^2)$ self-energy found to be concave down, while the corresponding IBMA  self energy is concave up.  It will be seen graphically that this change in the curvature at low energies becomes relevant in using SPA to determine the adsorption rate.

SPA assumes that the atom Green function has a simple pole in the lower-half of the complex energy plane that is located close to the real axis.  By linearizing the self-energy about an assumed quasiparticle energy $E_p$, one obtains that $E_p$ is determined by
\begin{equation}
E_p-E_k=\Sigma_r(E_p)
\label{QPenergy}
\end{equation}

The adsorption rate $\Gamma$ would then be given by
\begin{equation}
\Gamma\approx-2 Z\  \Sigma_i(E_p)
\label{rate}
\end{equation}
where the quasiparticle weight $Z=(1-{\partial\Sigma_r(E)\over\partial E}|_{E=E_p})^{-1}$.  

One can solve Eq.~\ref{QPenergy} graphically (see Fig.~\ref{fig:QP4}).  For sufficiently large $\epsilon$, I find there is a self-consistent solution to Eq.~\ref{QPenergy}; however,  with the large shift in the real part of the order $O(g_{kb}^2)$  self-energy $\Sigma_r$ at low $\epsilon$, I find that a self-consistent solution to Eq.~\ref{QPenergy} does not exist for $\epsilon\lesssim 0.4$ K (see Fig.~\ref{fig:QP}).  Thus, the SPA cannot be used to obtain an approximate adsorption rate in this case.  

In summary, the IBMA self-energy is a poor approximation to the exact self-energy to order $O(g_{kb}^2)$ for micrometer-sized samples of suspended graphene.  For $\epsilon\sim 1.7$ K, the magnitude of terms neglected in the IBMA self-energy exceeds  the magnitude of IBMA self-energy at zero energy. (This value of $\epsilon$ corresponds to a membrane radius $a\sim 0.07\mu$m.)

Using the order $O(g_{kb}^2)$ self-energy,  there are no solutions to Eq.~\ref{QPenergy} for the self-consistent quasiparticle energy $E_p$ for sufficiently small $\epsilon$.  Thus, the use of the SPA in this regime is invalid when using the order $O(g_{kb}^2)$ self-energy. The disagreement of the results of Ref.~\onlinecite{Sengupta2019} with previous work is a consequence of the use of the IBMA self-energy which does not contain the additional contributions essential to capturing the behavior at low $\epsilon$.

This work received support under NASA grant number 80NSSC19M0143.  The hospitality of JILA and the partial support of a JILA Visiting Fellowship are also gratefully acknowledged.  

\bibliographystyle{apsrev4-1}
\bibliography{comment}

\end{document}